\documentclass{aa}
\usepackage[varg]{txfonts}
\usepackage{graphicx}
\usepackage{natbib,twoopt}
\usepackage[breaklinks=true]{hyperref} 
\usepackage{lscape}
\usepackage{multirow}
\bibpunct{(}{)}{;}{a}{}{,} 
\makeatletter
 \newcommandtwoopt{\citeads}[3][][]{\href{https://ui.adsabs.harvard.edu/abs/#3/abstract}%
 {\def\hyper@linkstart##1##2{}%
 \let\hyper@linkend\@empty\citealp[#1][#2]{#3}}}
 \newcommandtwoopt{\citepads}[3][][]{\href{https://ui.adsabs.harvard.edu/abs/#3/abstract}%
 {\def\hyper@linkstart##1##2{}%
 \let\hyper@linkend\@empty\citep[#1][#2]{#3}}}
 \newcommandtwoopt{\citetads}[3][][]{\href{https://ui.adsabs.harvard.edu/abs/#3/abstract}%
 {\def\hyper@linkstart##1##2{}%
 \let\hyper@linkend\@empty\citet[#1][#2]{#3}}}
 \newcommandtwoopt{\citeyearads}[3][][]%
 {\href{https://ui.adsabs.harvard.edu/abs/#3/abstract}
 {\def\hyper@linkstart##1##2{}%
 \let\hyper@linkend\@empty\citeyear[#1][#2]{#3}}}
\makeatother

\begin{document}
   \title{Ejected from home: C/1980 E1 (Bowell) and C/2024 L5 (ATLAS)}
   \author{R. de la Fuente Marcos$^{1}$
           \and
           C. de la Fuente Marcos$^{2}$
           \and
           S.~J. Aarseth$^3$}
   \authorrunning{R. de la Fuente Marcos \and C. de la Fuente Marcos \and S.~J. Aarseth}
   \titlerunning{Ejected from the Solar System}
   \offprints{R. de la Fuente Marcos, \email{rauldelafuentemarcos@ucm.es}}
   \institute{$^{1}$AEGORA Research Group,
              Facultad de Ciencias Matem\'aticas,
              Universidad Complutense de Madrid,
              Ciudad Universitaria, E-28040 Madrid, Spain \\
              $^{2}$ Universidad Complutense de Madrid,
              Ciudad Universitaria, E-28040 Madrid, Spain \\
              $^3$Institute of Astronomy, University of Cambridge,
              Madingley Road, Cambridge CB3 0HA, UK}
   \date{Received 19 August 2024 / Accepted 25 September 2024}

   \abstract
      {Natural interstellar objects do not form isolated in deep space, but 
       escape their natal planetary systems. Early removal from their home 
       star systems via close flybys with still-forming planets could be the 
       dominant ejection mechanism. However, dynamically evolved planetary 
       systems such as the Solar System may also be a significant source of 
       natural interstellar objects. 
       }
      {We studied the dynamical evolution of two unusual Solar System 
       hyperbolic comets, C/1980~E1 (Bowell) and C/2024~L5 (ATLAS), to 
       investigate the circumstances that led them to reach moderate Solar 
       System excess hyperbolic speeds.
       }
      {We used $N$-body simulations and statistical analyses to explore the 
       planetary encounters that led to the ejection of C/1980~E1 and 
       C/2024~L5, and studied their pre- and post-encounter trajectories. 
       }
      {We confirm that C/1980~E1 reached its present path into interstellar
       space after an encounter with Jupiter at 0.23~au on December~9, 1980. 
       C/2024~L5 was scattered out of the Solar System following a flyby to
       Saturn at 0.003~au on January 24, 2022. Integrations backward in time
       show that C/1980~E1 came from the inner Oort cloud but C/2024~L5 
       could be a former retrograde, inactive Centaur. The receding 
       velocities of C/1980~E1 and C/2024~L5 when entering interstellar 
       space will be 3.8 and 2.8~km~s$^{-1}$, moving towards Aries and 
       Triangulum, respectively. 
       }
      {Our results for two comets ejected from the Solar System indicate 
       that dynamically evolved planetary systems can be effective sources 
       of interstellar objects and provide constraints on their velocity
       distribution. 
       }

   \keywords{comets: general --
             comets: individual: C/1980~E1 (Bowell) --
             comets: individual: C/2024~L5 (ATLAS) --
             methods: data analysis --
             methods: numerical -- celestial mechanics
            }

   \maketitle

   \section{Introduction\label{Intro}}
      Natural and artificial objects can reach interstellar space from their home planetary systems via the gravitational slingshot mechanism by 
      which, under the right conditions, a small body, moving past a more massive one also in motion, can be accelerated via conservation of momentum 
      and energy (see, e.g., \citealt{1974ApJ...190..253S}). For natural objects, this process is far more probable within dynamically young environs 
      such as the star-forming regions where star clusters and stellar associations are born (see, e.g., \citealt{2006M&PSA..41.5391B,
      2015tyge.conf...99D,2020RSOS....701271P}). Inside them, scattering by young, still-forming planets (see, e.g., \citealt{1978Icar...34..173F,
      2006Icar..184...59B}) and passing stars (see, e.g., \citealt{2018MNRAS.479L..17P,2021A&A...651A..38P}) may trigger bursts of ejected debris. In 
      this context, natural interstellar objects are mostly a by-product of the violent and chaotic processes that lead to the formation of planetary 
      systems (see, e.g., \citealt{1990PASP..102..793S}). 

      By the time planet formation is over, essentially $< 100$~Myr, large amounts of the material originally present in a protoplanetary disk may 
      already have been ejected (see, e.g., fig.~1 in \citealt{2019ApJ...874L..34P,2021A&A...651A..38P}); a fraction of this debris could be in the 
      form of interstellar objects of sizes similar to those of present-day asteroids and comets in the Solar System \citep{1976Icar...27..123S,
      1989ApJ...346L.105M}. However, the end of planet formation and the settling of planetary systems and their host stars as part of the field 
      population do not suppress the gravitational slingshot mechanism but merely reduce its effectiveness and efficiency. Most field stars remain in 
      the main sequence for $> 1$~Gyr and the continuous operation of the gravitational slingshot mechanism over extended periods of time, albeit at a 
      much lower efficiency, can still generate an unfaltering stream of interstellar objects.

      Here, we study the dynamical evolution of two rare Solar System hyperbolic comets, C/1980~E1 (Bowell) and C/2024~L5 (ATLAS), to investigate the 
      circumstances that led them to reach moderate excess hyperbolic speeds with respect to the barycenter of the Solar System. This paper is 
      organized as follows. In Sect.~\ref{DatToo}, we provide information on the input data and methods used in our numerical investigation. In 
      Sect.~\ref{E1}, we study the present dynamical status, and the past and future evolution of comet C/1980~E1; those of comet C/2024~L5 are 
      considered in Sect.~\ref{L5}. We discuss our dynamical results in Sect.~\ref{Dis} and summarize our conclusions in Sect.~\ref{Con}.  

   \section{Data and tools\label{DatToo}}
      To investigate the dynamical evolution of the objects under study here, we analyzed results from direct $N$-body simulations. To carry out these 
      calculations, we used orbit determinations and other relevant Solar System data (as of September 19, 2024) from the Jet Propulsion Laboratory's 
      (JPL) Small-Body Database (SBDB)\footnote{\href{https://ssd.jpl.nasa.gov/tools/sbdb\_lookup.html\#/}
      {https://ssd.jpl.nasa.gov/tools/sbdb\_lookup.html\#/}} provided by the Solar System Dynamics Group (SSDG, \citealt{2011jsrs.conf...87G,
      2015IAUGA..2256293G}).\footnote{\href{https://ssd.jpl.nasa.gov/}{https://ssd.jpl.nasa.gov/}} and input data from JPL's 
      \textsc{horizons}\footnote{\href{https://ssd.jpl.nasa.gov/?horizons}{https://ssd.jpl.nasa.gov/?horizons}} on-line solar system data and 
      ephemeris computation service, updated with the DE440/441 solution \citep{2021AJ....161..105P}. Data was retrieved from SBDB using the 
      \textsc{Python} package \textsc{Astroquery} \citep{2019AJ....157...98G} and its 
      \textsc{HorizonsClass}\footnote{\href{https://astroquery.readthedocs.io/en/latest/jplhorizons/jplhorizons.html}
      {https://astroquery.readthedocs.io/en/latest/jplhorizons/jplhorizons.html}} and
      \textsc{SBDBClass}\footnote{\href{https://astroquery.readthedocs.io/en/latest/jplsbdb/jplsbdb.html}
      {https://astroquery.readthedocs.io/en/latest/jplsbdb/jplsbdb.html}} classes.
%
%
      \begin{table*}
         \centering
         \fontsize{8}{11pt}\selectfont
         \tabcolsep 0.15truecm
         \caption{\label{elements}Keplerian orbital elements of comets C/1980~E1 (Bowell) and C/2024~L5 (ATLAS). 
                 }
         \begin{tabular}{lccccc}
            \hline\hline
            \multirow{2}{*}{Parameter}                         &   & \multicolumn{2}{c}{C/1980~E1 (Bowell)} & \multicolumn{2}{c}{C/2024~L5 (ATLAS)} \\ 
                                                               &   &  heliocentric           & barycentric  & heliocentric         & barycentric    \\
            \hline
             Perihelion, $q$ (au)                              & = &   3.363940$\pm$0.000003 &   3.355370   &   3.4320$\pm$0.0006  &   3.4400       \\
             Eccentricity, $e$                                 & = &   1.057733$\pm$0.000008 &   1.047673   &   1.0375$\pm$0.0003  &   1.0352       \\
             Inclination, $i$ (\degr)                          & = &   1.66174$\pm$0.00006   &   1.66170    & 166.5729$\pm$0.0003  & 166.6027       \\
             Longitude of the ascending node, $\Omega$ (\degr) & = & 114.557$\pm$0.002       & 114.5621     & 139.166$\pm$0.003    & 139.158        \\
             Argument of perihelion, $\omega$ (\degr)          & = & 135.083$\pm$0.002       & 135.1432     & 290.52$\pm$0.02      & 290.49         \\
             Mean anomaly, $M$ (\degr)                         & = & 359.84868$\pm$0.00003   & 359.88609    & 359.730$\pm$0.003    & 359.755        \\
             MOID with Jupiter (au)                            & = &   0.0108537             &   ---        &   0.00499388         &  ---           \\
             Total magnitude, $M_1$ (mag)                      & = &   5.8$\pm$1.0           &   ---        &   6.6$\pm$0.5        &  ---           \\
            \hline
         \end{tabular}
         \tablefoot{Values include the 1$\sigma$ uncertainty. The orbit of C/1980~E1 (solution date, April 15, 2021, 23:29:29 PST) is referred to 
                    epoch JD 2444972.5, which corresponds to 00:00 on 1982 January 3 TDB (Barycentric Dynamical Time, J2000.0 ecliptic and equinox) 
                    and it is based on 187 observations with a data-arc span of 2514~d. The orbit of C/2024~L5 (solution date, September 11, 2024, 
                    13:42:21 PDT) is referred to epoch JD 2460504.5, which corresponds to 00:00 on 2024 July 13 TDB and it is based on 215 
                    observations with a data-arc span of 82~d. Source: JPL's SBDB.
                   }
      \end{table*}
%
%

      The $N$-body simulations needed to investigate the evolution of the objects discussed in this paper were carried out using a direct $N$-body 
      code described by \citet{2003gnbs.book.....A} that is publicly available from the web site of the Institute of Astronomy of the University of 
      Cambridge.\footnote{\href{http://www.ast.cam.ac.uk/~sverre/web/pages/nbody.htm}{http://www.ast.cam.ac.uk/$\sim$sverre/web/pages/nbody.htm}} This 
      software makes use of the Hermite numerical integration scheme developed by \citet{1991ApJ...369..200M}. Our calculations do not include the 
      Galactic potential as they consist of integrations on comparatively short timescales (1~Myr), while the Sun takes $\sim$220~Myr to complete one 
      revolution around the center of the Galaxy. The orbit determinations in Table~\ref{elements} did not require non-gravitational terms to fit the 
      data; therefore, any contribution due to asymmetric outgassing is probably a second order effect in these cases and the impact of 
      non-gravitational forces could be safely neglected and was not included in the calculations. Additional technical details, relevant results from 
      this code as well as comparisons with results from other codes used for validation were presented in \citet{2012MNRAS.427..728D}. Our physical 
      model included the perturbations by the eight major planets, the Moon, the barycenter of the Pluto-Charon system, and the three largest 
      asteroids, Ceres, Pallas, and Vesta. Initial conditions for the comets under study here were generated by applying the Monte Carlo using the 
      Covariance Matrix (MCCM) methodology described in \citet{2015MNRAS.453.1288D}. The relevant covariance matrices were retrieved from JPL's SBDB 
      by using the {\tt SBDBClass} class. 
 
      Figures were produced using {\tt Matplotlib} \citep{2007CSE.....9...90H} and statistical tools provided by {\tt NumPy} 
      \citep{2011CSE....13b..22V,2020Natur.585..357H}. Histograms have statistically meaningful bin sizes and show the probability density so that the 
      area under the histogram integrates to 1. The bin width was computed using the Freedman-Diaconis rule \citep{FD81}. The Galactic space 
      velocities were computed as described by \citet{1987AJ.....93..864J} using the values of the relevant parameters provided by 
      \citet{2010MNRAS.403.1829S}.
%
%
      \begin{table*}
         \centering
         \fontsize{8}{11pt}\selectfont
         \tabcolsep 0.15truecm
         \caption{\label{orielements}Computed pre-planetary encounter Keplerian orbital elements of comets C/1980~E1 (Bowell) and C/2024~L5 (ATLAS). 
                 }
         \begin{tabular}{lccc}
            \hline\hline
             Parameter                                         &   & C/1980~E1 (Bowell)      & C/2024~L5 (ATLAS)        \\
            \hline
             Perihelion, $q$ (au)                              & = &   3.179958$\pm$0.000005 &   8.0$_{-0.4}^{+0.5}$    \\
             Eccentricity, $e$                                 & = &   0.999962$\pm$0.000008 &   0.70$\pm$0.02          \\
             Inclination, $i$ (\degr)                          & = &   1.77204$\pm$0.00010   & 153$\pm$2                \\
             Longitude of the ascending node, $\Omega$ (\degr) & = & 120.638$\pm$0.002       & 137.1$\pm$0.2            \\
             Argument of perihelion, $\omega$ (\degr)          & = & 134.485$\pm$0.002       & 240$_{-8}^{+6}$          \\
            \hline
         \end{tabular}
         \tablefoot{The pre-planetary encounter orbit estimate of C/1980~E1 includes mean values and 1$\sigma$ uncertainties and it is the result of
                    10$^{3}$ $N$-body calculations back in time as described in the text. The one of C/2024~L5 shows medians and 16th and 84th 
                    percentiles from 10$^{4}$ simulations. 
                   }
      \end{table*}
%
%

   \section{Ejected by Jupiter: C/1980~E1 (Bowell)\label{E1}}
      Comet C/1980~E1 (Bowell) --- also known as Comet~Bowell~1982~I~=~1980b --- was first imaged by E.~L.~G.~Bowell on February 11, 1980, while 
      observing at the Anderson Mesa Station of the Lowell Observatory in Arizona \citep{1980IAUC.3461....1B} with the 33-cm astrograph (the same 
      instrument used to discover Pluto 50 years before), although the actual discovery image was acquired on March 13, 1980, with the same telescope. 
      When first observed, it was moving inwards at 7.24~au from the Sun and was described as diffuse, with no obvious condensation. The comet reached 
      perihelion on March 12, 1982, and it was last observed on December 30, 1986, from the Steward Observatory in Kitt Peak by the Spacewatch project 
      0.9-m telescope, when it was located at 13.92~au from the Sun, heading for the outskirts of the Solar System.

      The orbit determination of C/1980~E1 available from JPL's SBDB and computed on April 15, 2021 (see Table~\ref{elements}), is based on 187 data 
      points for a data-arc span of 2514~d, and is referred to epoch JD 2444972.5 that is the instant $t = 0$ for these calculations. With a 
      current value of the heliocentric orbital eccentricity, $e$, of 1.0577 and a barycentric one of 1.0477 --- fourth only to those of 1I/2017~U1 
      (`Oumuamua), the first known interstellar object (see, e.g.,  \citealt{2017MPEC....V...17W,2017Natur.552..378M,2018Msngr.173...13H,
      2019NatAs...3..594O}), 2I/Borisov, the first {\it bona fide} interstellar comet (see, e.g., \citealt{2019RNAAS...3..131D,2019ApJ...885L...9F,
      2019ApJ...886L..29J,2020NatAs...4...53G}), and the poorly-known comet C/1954~O1 (Vozarova) --- it has the lowest value of the orbital 
      inclination (1\fdg66174) among those of known nearly-parabolic ($e$$\sim$1) to hyperbolic ($e>1$) small bodies. The cause of its present high 
      eccentricity is well understood --- it was the result of a close encounter with Jupiter on December 9, 1980 (see, e.g., 
      \citealt{1980IAUC.3465....1B,1982M&P....26..311B,2013RMxAA..49..111B}) --- but its actual origin and pre-discovery orbital evolution are still 
      unclear. 

      Soon after discovery, it was argued that C/1980~E1 could be a first-time visitor from the Oort cloud \citep{1980IAUC.3468....1S}, recently 
      perturbed by a stellar flyby or perhaps an interstellar comet \citep{1981MNRAS.196P..45H}. \citet{2017AJ....153..133E} pointed out that the 
      orbital properties of C/1980~E1 are compatible with those of interstellar objects although bordering those of slightly perturbed Oort cloud 
      comets, but it is still widely assumed that the origin of this comet is in the Oort cloud (see, e.g., \citealt{2017MNRAS.472.4634K,
      2018AJ....156...73H}), although no detailed calculations have been published since the latest public orbit determination was announced. With a 
      data-arc span of 6.88~yr and being hyperbolic at nearly the 5990$\sigma$ level (barycentric), a detailed numerical exploration using the latest 
      data may help to confirm the dynamical status of C/1980~E1 prior to discovery and its close planetary encounter.

      We explored the close encounter of C/1980~E1 with Jupiter by performing 10$^{3}$ $N$-body simulations based on the MCCM methodology as 
      pointed out in Sect.~\ref{DatToo}. The median and 16th and 84th percentiles of the minimum approach distance to Jupiter during the flyby 
      matched their average value and its standard deviation, as expected from the small uncertainties of the orbit determination (see 
      Table~\ref{elements}). The minimum approach distance was 0.228122$\pm$0.000006~au, the Hill radius of Jupiter is 0.338~au, and the calendar 
      date for this flyby was 1980-Dec-09 11:06; \textsc{horizons} gives a value of 1980-Dec-09 11:03$\pm$00:09 and a nominal minimum approach 
      distance to Jupiter of 0.22841~au. Table~\ref{orielements} shows the pre-encounter orbit of C/1980~E1 as reconstructed from the results of 
      our calculations. It corresponds to a near-parabolic comet with an orbital eccentricity (median and 16th and 84th percentiles) of 
      0.999961$_{-0.000007}^{+0.000008}$. Although C/1980~E1 now follows a hyperbolic path, its orbit prior to encountering Jupiter was bound and
      similar to those of thousands of known Solar System comets. Our analysis of short-term integrations assigns exactly zero probability to the 
      hypothesis of having a hyperbolic orbit right before its encounter with Jupiter. 

\begin{figure}
        \centering
         \includegraphics[width=\linewidth]{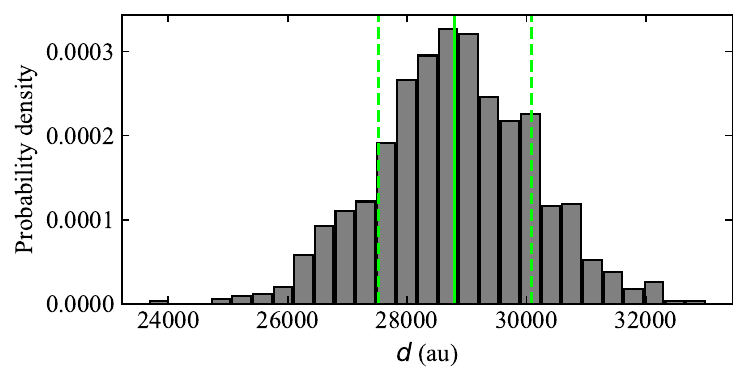}
         \includegraphics[width=\linewidth]{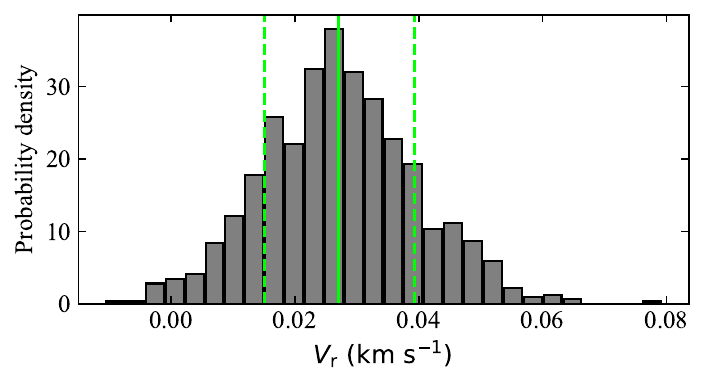}
         \caption{Range or distance and range rate or radial velocity of C/1980~E1 (Bowell), 1.0~Myr before its close encounter with Jupiter. 
                  {\it Top panel:} Distribution of distances at the end of the simulations, 28792$_{-1272}^{+1288}$~au. 
                  {\it Bottom panel:} Distribution of radial velocities at the end of the calculations, 0.027$\pm$0.012~km~s$^{-1}$.
                  Distributions resulting from the evolution of 10$^{3}$ control orbits. The median is displayed as a continuous vertical green 
                  line, the 16th and 84th percentiles as dashed lines.
                 }
         \label{preE1}
      \end{figure}

      However, its minimum orbit intersection distance (MOID) with Jupiter was sufficiently small to have led to close flybys with Jupiter in 
      previous passages through the inner Solar System. Within this dynamical context, C/1980~E1 might be an old comet, a new one recently 
      dislodged from the Oort cloud or, less likely, a low-relative-velocity, temporary capture from interstellar space --- under the right 
      circumstances, the gravitational slingshot mechanism can lead to a capture instead of an ejection. The analysis of the evolution into the 
      past of 10$^{3}$ control orbits of C/1980~E1 generated using an MCCM process shows that 1~Myr ago, C/1980~E1 was located at 
      28792$_{-1272}^{+1288}$~au from the Sun, moving with a radial velocity of 0.027$\pm$0.012~km~s$^{-1}$ (see Fig.~\ref{preE1}). 
      \citet{1982ApJ...255..307V} showed that a relative velocity exceeding 0.5~km~s$^{-1}$ when near the Hill radius of the Solar System is required 
      to become interstellar. An interstellar origin, as a recent capture, is therefore strongly rejected.
%
%
%
%

\begin{figure}
        \centering
         \includegraphics[width=\linewidth]{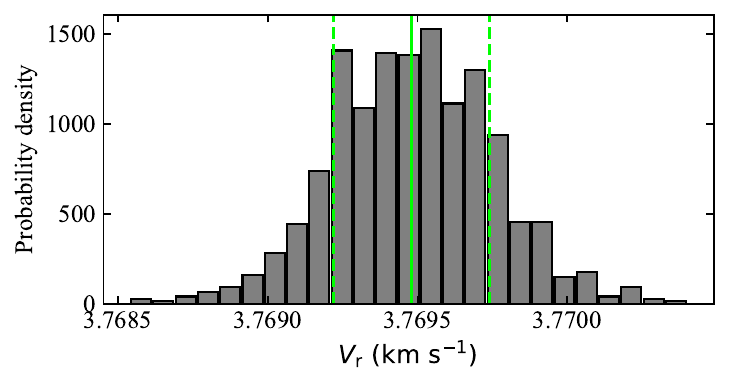}
         \caption{Receding velocity of C/1980~E1 (Bowell) after escaping the Solar System. Distribution resulting from the evolution of 10$^{3}$ 
                  control orbits for 1.0~Myr. The median is displayed as a continuous vertical green line, the 16th and 84th percentiles as dashed 
                  lines, 3.7695$\pm$0.0003~km~s$^{-1}$.
                 }
         \label{apexE1}
      \end{figure}

\begin{figure}
        \centering
         \includegraphics[width=\linewidth]{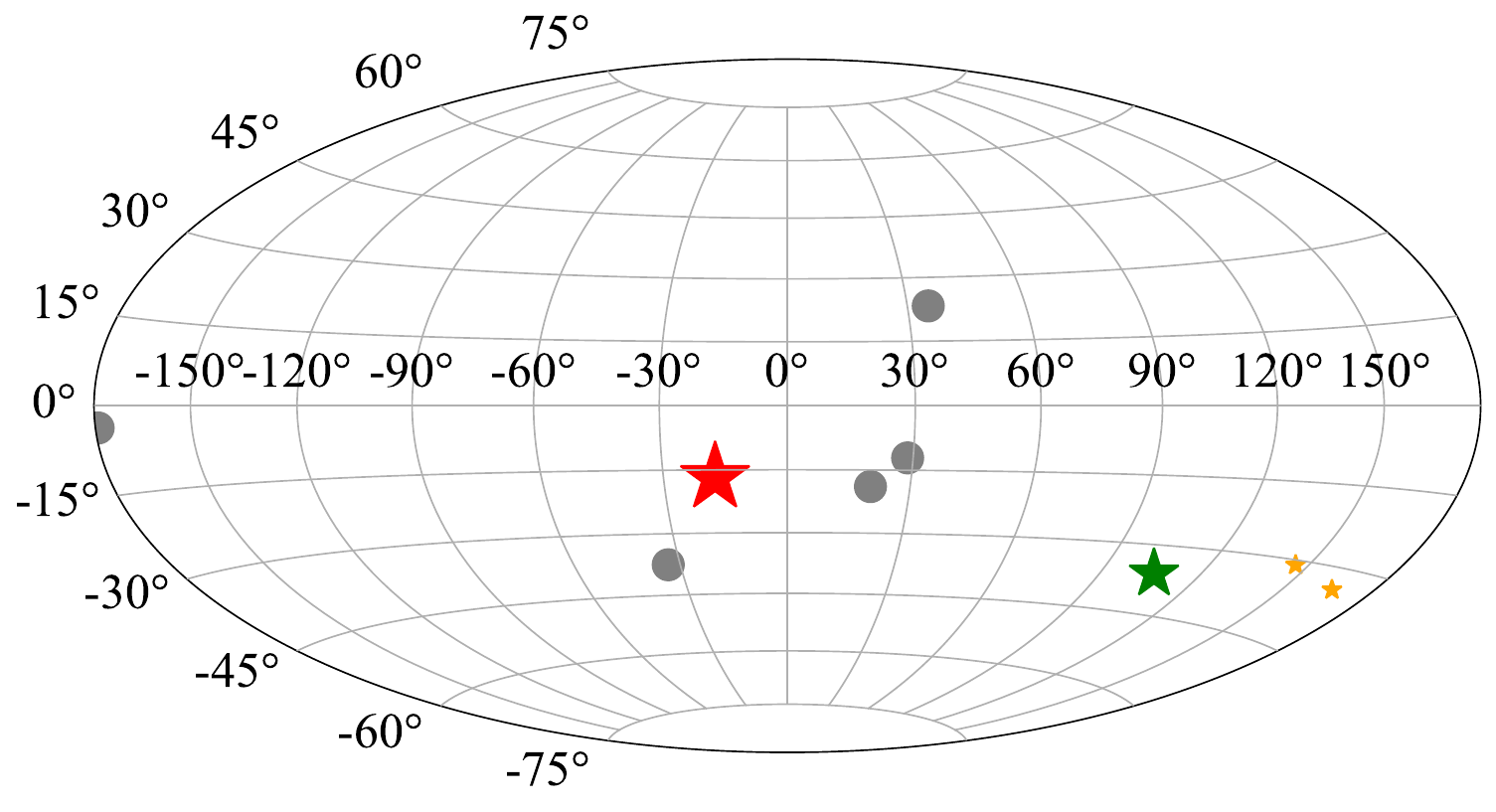}
         \includegraphics[width=\linewidth]{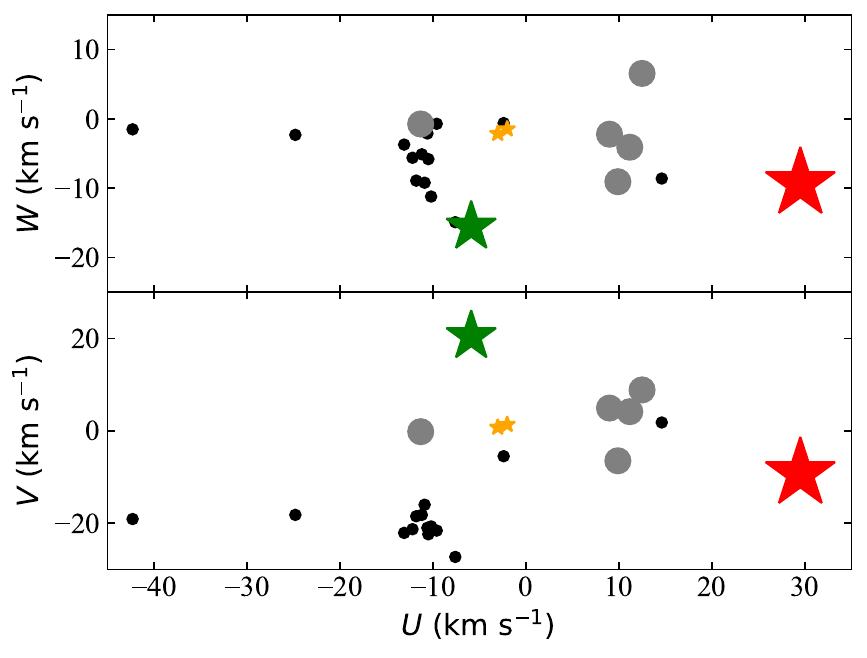}
         \caption{Apex and velocity of known natural and artificial interstellar objects. 
                  {\it Top panel:} Location of the apex of each interstellar object in Galactic coordinates. Comet 2I/Borisov is shown as a red star 
                  and 1I/2017~U1 (`Oumuamua) is displayed in green. From left to right, C/2024 L5 (ATLAS) and C/1980 E1 (Bowell) are displayed as 
                  orange stars, Pioneer~10, Voyager~2, New~Horizons, Pioneer~11, and Voyager~1 are shown as grey circles. Coordinates are displayed in 
                  a Hammer-Aitoff equal-area projection. {\it Bottom panels:} Heliocentric Galactic velocity components at apex of the same objects 
                  and velocities of stellar groups within 100~pc from the Sun in black.
                 }
         \label{apexes}
      \end{figure}

      On the other hand, from a similar set of calculations forward in time, at 3.8595$\pm$0.0003~pc from the Sun and 1.0~Myr into the future, 
      C/1980~E1 will be receding from us (see Fig.~\ref{apexE1}) at 3.7695$\pm$0.0003~km~s$^{-1}$ towards (apex)
      $\alpha$=03$^{\rm h}~16^{\rm m}~34.9^{\rm s}$, $\delta$=$+$16\degr~37\arcmin~00.1{\arcsec} (49\fdg1454$\pm$0\fdg0012, 
      $+$16\fdg6167$\pm$0\fdg0003) in the constellation of Aries with Galactic coordinates $l$=166\fdg00, $b$=$-$33\fdg85 (see Fig.~\ref{apexes}, top
      panel), and ecliptic coordinates $\lambda$=51\fdg17, $\beta$=$-$01\fdg49. The components of its heliocentric Galactic velocity will be $(U, V, 
      W)$=$(-$3.0392$\pm$0.0002, $+$0.7532$\pm$0.0001, $-$2.0988$\pm$0.0002)~km~s$^{-1}$ (see Fig.~\ref{apexes}, bottom panels). 
%
%
%
%
%
%
%
%

   \section{Ejected by Saturn: C/2024~L5 (ATLAS)\label{L5}}
      Comet C/2024~L5 (ATLAS) was discovered on June 14, 2024, as A117uUD, by the Asteroid Terrestrial-impact Last Alert System (ATLAS, 
      \citealt{2018PASP..130f4505T}) observing with the unit located in the Sutherland plateau in South Africa \citep{2024MPEC....O...15R,
      2024CBET.5418....1G}. The orbit determination of C/2024~L5 available from JPL's SBDB and computed on September 11, 2024 (see 
      Table~\ref{elements}), is based on 215 data points for a data-arc span of 82~d, and is referred to epoch JD 2460504.5 that is the instant 
      $t=0$ for these calculations. With a current value of the heliocentric orbital eccentricity, $e$, of 1.0375 and a barycentric one of 1.0352, 
      C/2024~L5 follows C/1980~E1 (Bowell) as the known small body with the fifth highest value of $e$. Based on early data, S.~Nakano 
      \citep{2024CBET.5418....1G} and \citet{2024RNAAS...8..184D} concluded that the comet had experienced a very close encounter with Saturn on 
      January 24, 2024, leading to its present hyperbolic trajectory ($\sim$126$\sigma$ level, barycentric) from an initially bound ($e<1$) but 
      retrograde ($i>90\degr$) orbit.

\begin{figure}
        \centering
         \includegraphics[width=\linewidth]{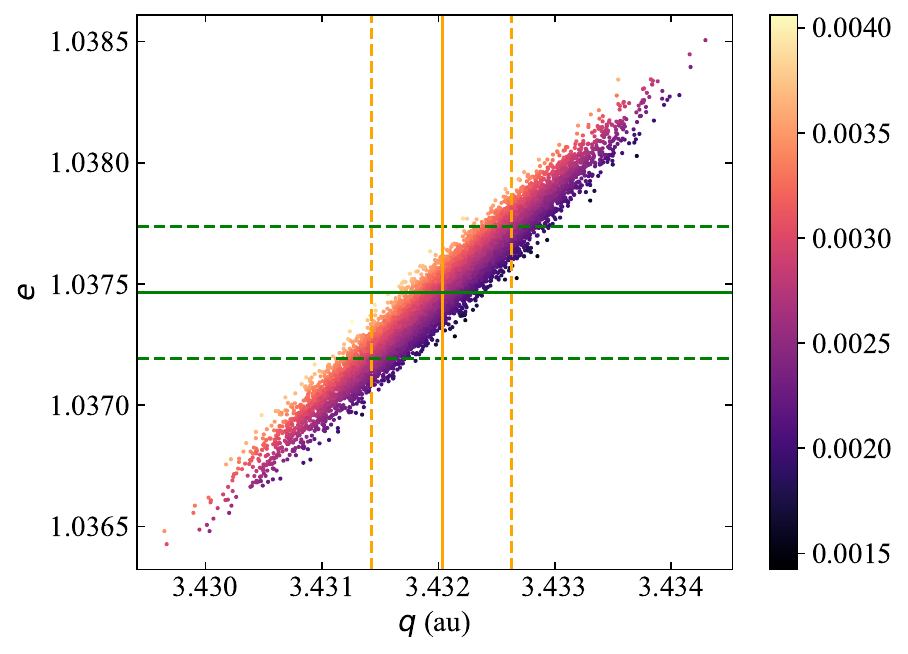}
         \includegraphics[width=\linewidth]{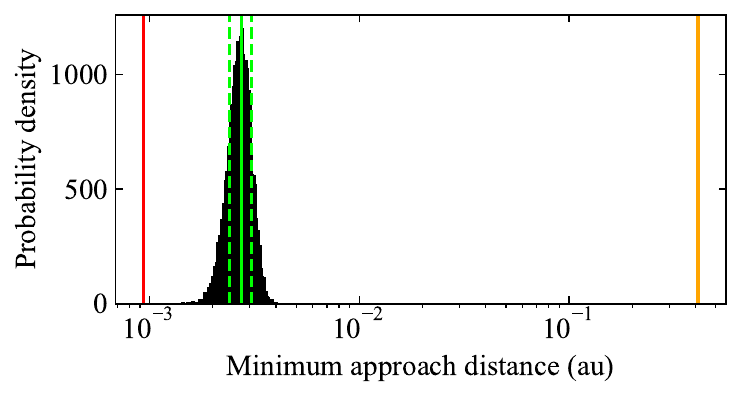}
         \includegraphics[width=\linewidth]{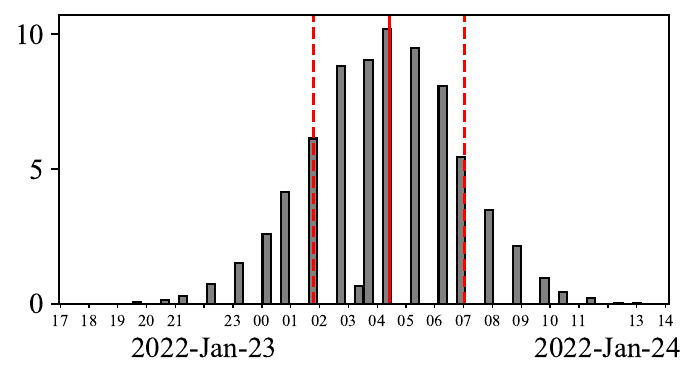}
         \caption{Close encounter of C/2024~L5 (ATLAS) with Saturn. {\it Top panel:} Minimum approach distance (color-coded) as a function of the 
                  values of ($q$, $e$) at $t=0$. Median values are shown as continuous (green for $e$, orange for $q$) lines, 16th and 84th 
                  percentiles as dashed lines. {\it Middle panel:} Distribution of minimum approach distances, the median and 16th and 84th 
                  percentiles, 0.0027$\pm$0.0003~au, are shown in green, the Hill radius of Saturn is 0.412~au (in orange), and the Roche radius is 
                  0.000932~au (in red). {\it Bottom panel:} Distribution of calendar dates for the flyby, the median and 16th and 84th percentiles (in 
                  red) are 2022-Jan-24 04:25$\pm$02:38. The output cadence (time resolution) in our calculations was 0.877~h. 
                 }
         \label{flybydistance}
      \end{figure}

\begin{figure}
        \centering
         \includegraphics[width=\linewidth]{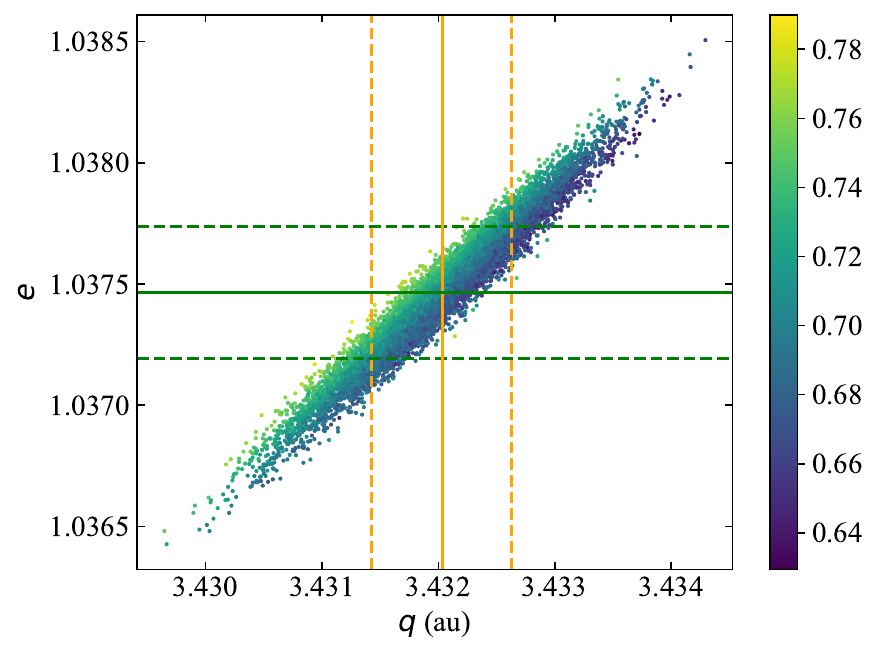}
         \includegraphics[width=\linewidth]{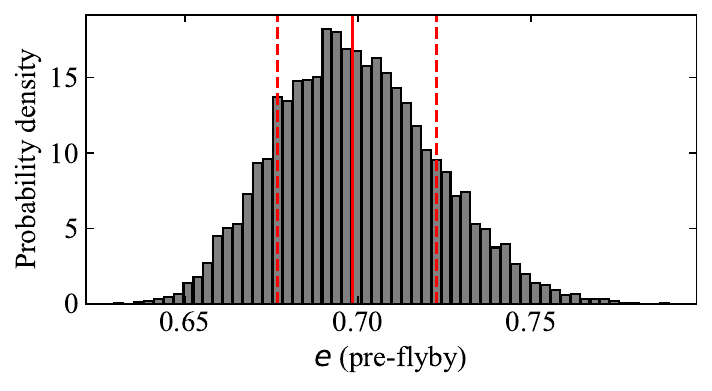}
         \caption{Distribution of the pre-flyby eccentricity of C/2024~L5 (ATLAS). {\it Top panel:} Values of the pre-flyby eccentricity (color-coded) 
                  as a function of the input values of ($q$, $e$) at $t=0$. Median values are shown as continuous (green for $e$, orange for $q$)
                  lines, 16th and 84th percentiles as dashed lines. {\it Bottom panel:} Distribution of the computed pre-flyby eccentricity. The median 
                  is displayed as a continuous red line, 16th and 84th percentiles as dashed lines, 0.70$\pm$0.02.
                 }
         \label{preflybyecc}
      \end{figure}

      Using the latest public orbit determination in Table~\ref{elements}, we studied the close encounter of C/2024~L5 with Saturn by performing 
      10$^{4}$ $N$-body simulations based on the MCCM methodology (see Sect.~\ref{DatToo}). Our results are summarized in Figs.~\ref{flybydistance} 
      and \ref{preflybyecc}. The comet experienced an encounter with Saturn at very close range in 2022 and this makes the reconstruction of its 
      pre-encounter path difficult. The top panel in Fig.~\ref{flybydistance} displays the minimum approach distance (color-coded) to Saturn during 
      the flyby as a function of the ($q$, $e$) values for each synthetic orbit. The median and 16th and 84th percentiles, 0.0027$\pm$0.0003~au, 
      indicate that the encounter took place well inside the Hill radius of Saturn, 0.412~au, but not inside the Roche radius (see, e.g., 
      \citealt{2014IAUS..310..182C}) of the planet, 0.000932~au (see Fig.~\ref{flybydistance}, middle panel). Therefore, it is highly unlikely that 
      its current level of cometary activity could be the result of any flyby-induced fragmentation event. Consistently, the comet was inactive by 
      mid-2023 (S.~Deen 2024, private communication). The histogram of calendar dates for this flyby in Fig.~\ref{flybydistance}, bottom panel, shows 
      a median of 2022-Jan-24 04:25, \textsc{horizons} gives 2022-Jan-24 02:51$\pm$07:43 and a nominal minimum approach distance to Saturn of 
      0.00269~au. 

%
%
%
%
%
%
%
%

\begin{figure}
        \centering
         \includegraphics[width=\linewidth]{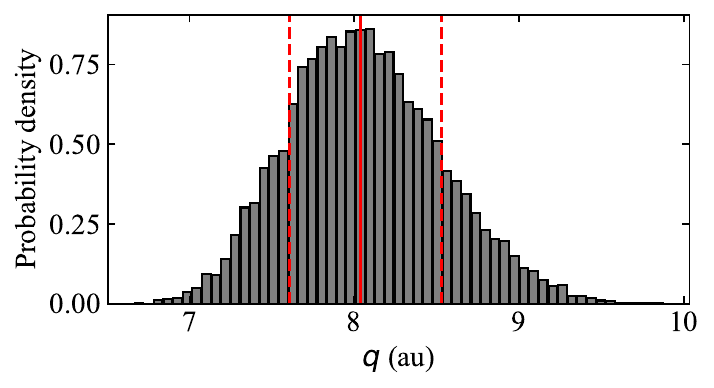}
         \includegraphics[width=\linewidth]{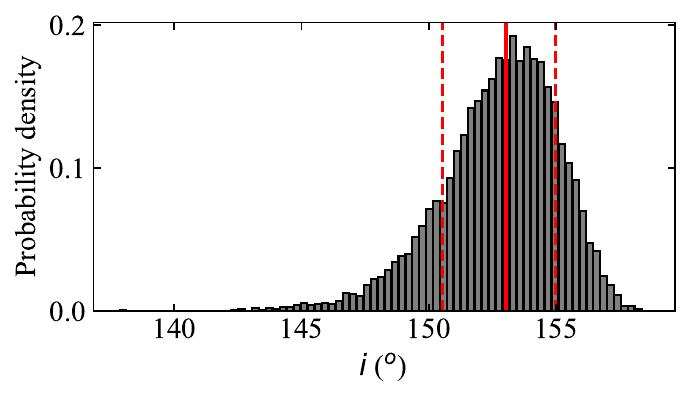}
         \includegraphics[width=\linewidth]{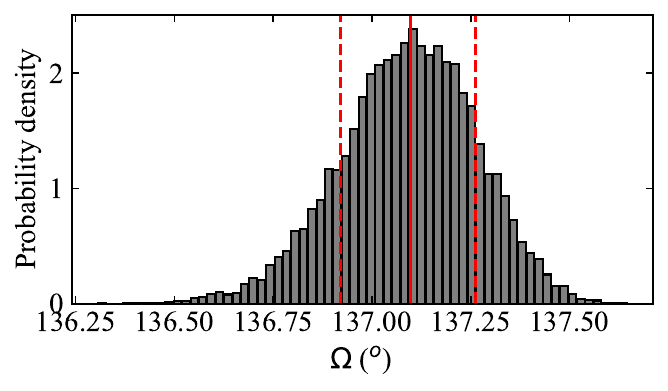}
         \includegraphics[width=\linewidth]{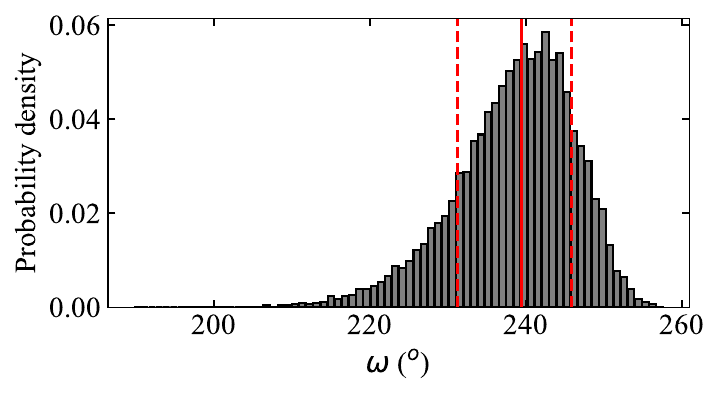}
         \caption{Additional orbital elements of the pre-flyby orbit of C/2024~L5 (ATLAS). {\it Top panel:} Distribution of the computed pre-flyby 
                  perihelion distance. {\it Second to top panel:} Orbital inclination. {\it Second to bottom panel:} Longitude of the ascending node.
                  {\it Bottom panel:} Argument of perihelion. Median values are displayed as continuous red lines, 16th and 84th percentiles as dashed 
                  lines.
                 }
         \label{preflybyorb}
      \end{figure}

      As for the dynamical nature of this comet prior to its encounter with Saturn, Fig.~\ref{preflybyecc} shows the distribution of the resulting 
      pre-flyby eccentricity; the top panel displays the color-coded values as a function of ($q$, $e$); the bottom panel presents the histogram of  
      pre-encounter eccentricities with median and 16th and 84th percentiles of 0.70$\pm$0.02. The probability of C/2024~L5 having a 
      pre-encounter orbital eccentricity equal to or higher than that of `Oumuamua, 1.20113, or 2I/Borisov, 3.35648, is 0.0; the probability of having 
      a non-hyperbolic pre-encounter trajectory is 1.0. Our best estimate for the orbit of C/2024~L5 prior to its encounter with Saturn is shown in 
      Table~\ref{orielements} and in Figs.~\ref{preflybyecc} and \ref{preflybyorb}. 
%
%
%
%

\begin{figure}
        \centering
         \includegraphics[width=\linewidth]{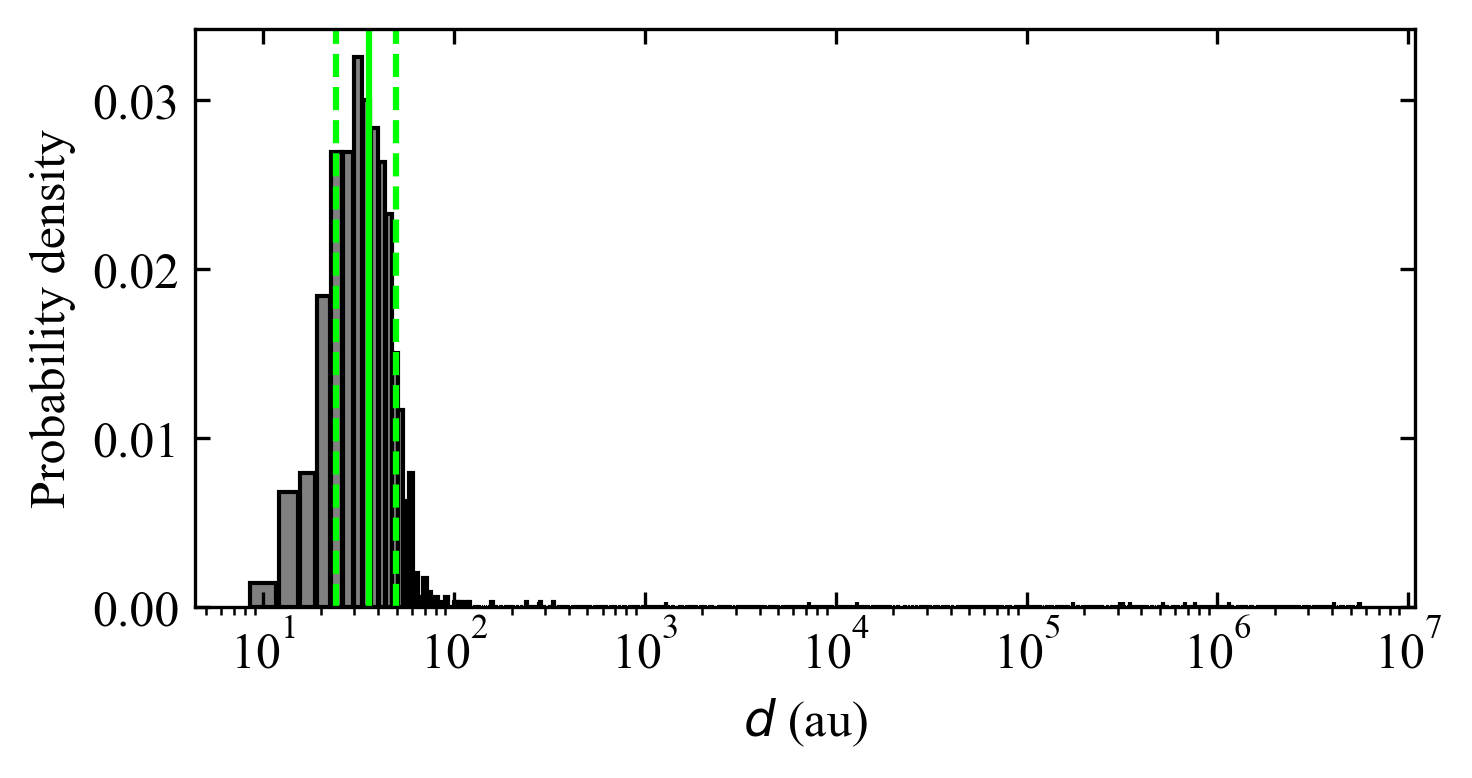}
         \includegraphics[width=\linewidth]{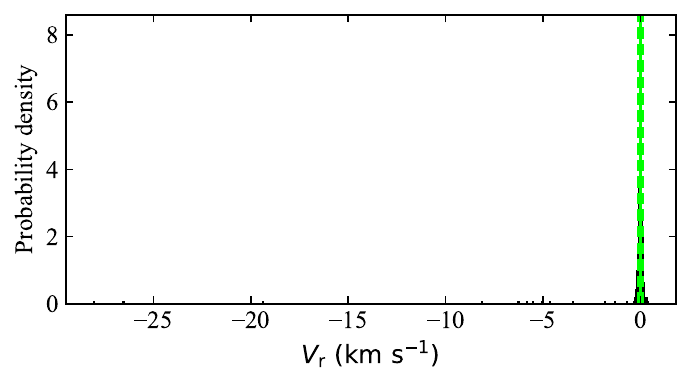}
         \caption{Range or distance and range rate or radial velocity of C/2024~L5 (ATLAS), 1.0~Myr prior to its encounter with Saturn. 
                  {\it Top panel:} Distribution of distances at the end of the simulations, 36$_{-12}^{+14}$~au. 
                  {\it Bottom panel:} Distribution of radial velocities at the end of the calculations, 0.00$_{-0.09}^{+0.08}$~km~s$^{-1}$.
                  Distributions resulting from the evolution of 10$^{3}$ control orbits. The median is displayed as a continuous vertical green 
                  line, the 16th and 84th percentiles as dashed lines.
                 }
         \label{preL5}
      \end{figure}

      Having low MOID with both Jupiter and Saturn, this object may have had a very chaotic dynamical past. The analysis of the evolution of 10$^{3}$ 
      control orbits of C/2024~L5 generated using an MCCM process and integrated backward in time shows that 1.0~Myr ago, C/2024~L5 was located at 
      36$_{-12}^{+14}$~au from the Sun, so around the trans-Neptunian region, moving with a radial velocity of 0.00$_{-0.09}^{+0.08}$~km~s$^{-1}$ (see 
      Fig.~\ref{preL5}). An interstellar origin (as a recent capture) is strongly rejected, the probability of capture from interstellar space is 
      0.00$_{-0.00}^{+0.02}$ for integrations 1.0~Myr into the past (several sets). 
%
%
%
%

\begin{figure}
        \centering
         \includegraphics[width=\linewidth]{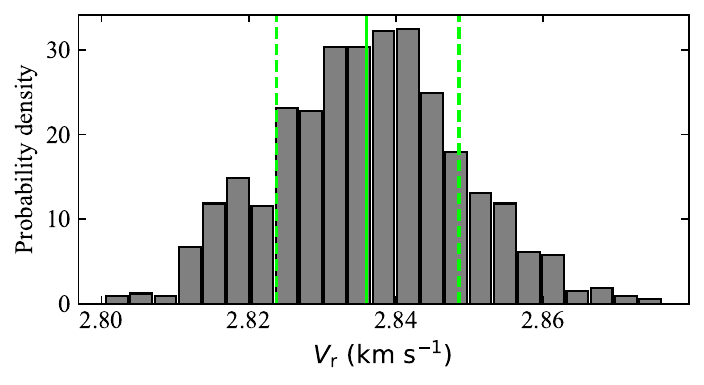}
         \caption{Receding velocity of C/2024~L5 (ATLAS) after escaping the Solar System. Distribution resulting from the evolution of 10$^{3}$ 
                  control orbits for 1.0~Myr. The median is displayed as a continuous vertical green line, the 16th and 84th percentiles as 
                  dashed lines, 2.836$_{-0.012}^{+0.013}$~km~s$^{-1}$.
                 }
         \label{apexL5}
      \end{figure}

      Regarding its future, the analysis of the evolution of a similar set of 10$^{3}$ control orbits shows that all of them lead to escaping from 
      the Solar System with a velocity of 2.836$_{-0.012}^{+0.013}$km~s$^{-1}$ after 1.0~Myr (see Fig.~\ref{apexL5}), when the comet will be located 
      2.906$\pm$0.013~pc from the Solar System receding from us towards $\alpha$=02$^{\rm h}~21^{\rm m}~14^{\rm s}$, 
      $\delta$=$+$28\degr~00\arcmin~00{\arcsec} (35\fdg31$\pm$0\fdg09, $+$28\fdg00$\pm$0\fdg03) in the constellation of Triangulum with Galactic 
      coordinates $l$=146\fdg05, $b$=$-$30\fdg81, and ecliptic coordinates $\lambda$=42\fdg27, $\beta$=$+$13\fdg16. The components of its 
      heliocentric Galactic velocity will be $(U, V, W)$=$(-$2.022$\pm$0.009, $+$1.358$\pm$0.006, $-$1.453$\pm$0.007)~km~s$^{-1}$ (see 
      Fig.~\ref{apexes}, bottom panels).
%
%
%
%

   \section{Discussion\label{Dis}}
      Our numerical results strongly reject an extrasolar origin for C/1980~E1 (Bowell), as a recent capture; for C/2024~L5 (ATLAS), it emerges as a 
      low-probability setting. Here, we focus on the probable dynamical classes of their parent bodies. Being part of a certain dynamical class at any 
      given time does not necessarily imply that a small body was formed in the region of the Solar System that is hosting that class. A dramatic 
      example of this fact is in the Manx comets that are made of materials originally formed in the inner Solar System although their present-day 
      trajectories come from the Oort cloud \citep{2016SciA....2E0038M}. Unfortunately, the numerical reconstruction of the past orbital evolution of 
      objects in highly-chaotic paths has limitations. The conservation of the Tisserand invariant can be used to identify some of the original 
      parameters of such orbits \citep{2022MNRAS.510..276N,2024MNRAS.527.4889N}. Spectroscopic observations (or in situ analyses via robotic missions) 
      can provide reliable information on the region where the material that constitutes the small body was originally processed (see, e.g., 
      \citealt{1994AREPS..22..457P}). 
      
      On the other hand, \citet{2021A&A...651A..38P} argued that the ejection velocity of interstellar objects produced by planet scattering is 
      $\sim$4--8~km~s$^{-1}$. Our calculations suggests either a smaller lower limit for the ejection velocity after planetary encounters in evolved 
      planetary systems, or a lower median or most probable value, as the receding velocities of C/1980~E1 and C/2024~L5 when entering interstellar 
      space will be 3.8 and 2.8~km~s$^{-1}$, respectively.

      In addition, here we provide a basic analysis of a different aspect of the problem of production of interstellar objects, that of the artificial 
      ones. Over four decades ago, when the Pioneer~10 probe reached the escape velocity, the Solar System started producing artificial interstellar 
      objects, 4.5682~Gyr after its own formation \citep{2010NatGe...3..637B}. Alien civilizations as technologically advanced as ours are expected to 
      produce artificial interstellar objects as well \citep{1960Natur.186..670B,1963P&SS...11..485S}. Our deep space probes escape the Solar System 
      after one or more gravity-assist maneuvers, which are precisely planned applications of the gravitational slingshot mechanism (see, e.g., 
      \citealt{1968JSpRo...5..633S,1968JSpRo...5.1029F}).

      \subsection{An inner Oort cloud origin for C/1980~E1}
         Our calculations (see Sect.\ref{E1}) place the origin of C/1980~E1 (Bowell) in the torus-shaped inner Oort cloud or Hills cloud 
         \citep{1981AJ.....86.1730H,2001AJ....121.2253L}, not the spherical classical Oort cloud whose outer edge might be 10$^{5}$~au from the Sun 
         \citep{1950BAN....11...91O}. Inner Oort cloud candidate members have been discovered during the last two decades \citep{2004ApJ...617..645B,
         2014Natur.507..471T,2019AJ....157..139S}. On the other hand, it has a spectrum consistent with an origin in the Solar System 
         \citep{1982AJ.....87.1854J}. Based on the available evidence, the hypothesis of an origin in interstellar space (as a recent capture) for 
         this comet can be discarded. 

      \subsection{A Centaur origin for C/2024~L5 (ATLAS)}
         As for the origin of C/2024~L5 (ATLAS), our calculations (see Sect.\ref{L5} and Table~\ref{orielements}) strongly suggest a Solar System 
         provenance.  However, JPL's SBDB includes no objects with values of $q$, $e$ and $i$ within the ranges shown in Table~\ref{orielements}. When 
         no constraint on the value of $i$ is considered, we find one entry in the database, 2017~GY$_{8}$, a prograde Centaur --- Centaurs are 
         objects with orbits between those of Jupiter and Neptune (5.5~au $<a<$ 30.1~au, $a$ is the semimajor axis) and $i<90\degr$. 
         \citet{2003AJ....126.3122T} found that nearly two thirds of the objects in Centaur orbits are expected to be eventually ejected into 
         interstellar space.

         Just outside the orbital domain outlined in Table~\ref{orielements}, we find 2017~UX$_{51}$ with $q$=7.61~au, $e$=0.75, and $i$=90\fdg45. 
         This dynamical context and our own numerical results strongly suggest that, prior to its close planetary encounter with Saturn, C/2024~L5 was 
         in a very unstable, retrograde trajectory similar to those of retrograde Centaurs. \citet{2013Icar..224...66V} argued that retrograde 
         Centaurs are unlikely to come from the trans-Neptunian or Kuiper belt. The analysis of the past orbital evolution of known retrograde 
         Centaurs in \citet{2014Ap&SS.352..409D} concluded that they may come from the Oort cloud but the existence of a closer, previously unknown 
         reservoir cannot be ruled out. \citet{2020MNRAS.494.2191N} argued for an interstellar origin of the high-inclination Centaurs, but as early
         captures that took place during the early stages of the formation of the Solar System. This scenario is consistent with other studies that 
         argue for the presence of a significant fraction of extrasolar debris --- originally captured from the surroundings of the nascent Solar 
         System --- in the present-day Oort cloud (see, e.g., \citealt{2010Sci...329..187L,2021A&A...652A.144P}). \citet{2018P&SS..158....6F} found 
         that retrograde Centaurs do not exhibit cometary activity and that their dynamical lifetimes are comparatively long. 
         \citet{2019A&A...630A..60L} pointed out that under certain conditions prograde Centaurs may turn retrograde and vice versa. Such orbital 
         flips can be triggered by the von~Zeipel--Lidov--Kozai mechanism \citep{1910AN....183..345V,1962P&SS....9..719L,1962AJ.....67..591K,
         2019MEEP....7....1I} via a secular resonance with, e.g., Jupiter \citep{2015MNRAS.446.1867D}. This dynamical pathway has been explored by 
         \citet{2020P&SS..19105031K} for 2017~UX$_{51}$. However, an orbital flip for C/2024~L5 is not favored by our 1.0~Myr integrations into the 
         past as all the control orbits remain retrograde. An interstellar origin as a recent capture seems improbable. Future astrometric and 
         spectroscopic observations may help to confirm its source region. This comet will reach perihelion on March 10, 2025. 

      \subsection{Leaving the Solar System: Natural versus artificial} 
         We have also placed artificial materials in hyperbolic trajectories --- the Pioneer~10 and 11 \citep{1993AdSpR..13..267D}, Voyager~1 and 2 
         \citep{1993AdSpR..13R.301S}, and New Horizons \citep{2010cosp...38..634W} deep space probes, and secondary hardware from these missions such 
         as rocket upper stages --- which may eventually be observed by extraterrestrial intelligences. 1I/2017~U1, 2I/Borisov, C/1980~E1 (Bowell), 
         C/2024~L5 (ATLAS), and the five deep space probes are on their way out of the Solar System and they are not coming back. For completeness, 
         we compare here the direction that these objects travel with respect to the Sun or apex, and also their barycentric excess hyperbolic speeds 
         and heliocentric Galactic velocities. By studying the properties of apexes and velocities, we might be able to understand what separates 
         natural intruders from artificial ones, dynamically. For this section, we have performed integrations forward in time for 10$^{5}$~yr of the 
         nominal orbits of the five space probes using the latest JPL's SBDB data. Our computations neglect the possible existence of anomalous 
         accelerations (see, e.g., \citealt{1998PhRvL..81.2858A}). Our calculations are updated versions of the ones in \citet{1997AcAau..40..383R} 
         and \citet{2019RNAAS...3...59B}.

         The apex of Pioneer~10 is located towards $\alpha=5^{\rm h}~33^{\rm m}~40^{\rm s}$ and $\delta=+26\degr~13\arcmin~1\arcsec$, in Taurus. Our
         calculations place the probe at 1.16~pc from the Sun receding from us at 11.4~km~s$^{-1}$. The components of its Galactic velocity will be 
         then $(U, V, W)$=$(-11.32, -0.19, -0.73)$~km~s$^{-1}$. For Pioneer~11, we found $\alpha=19^{\rm h}~27^{\rm m}~18^{\rm s}$ and
         $\delta=-9\degr~13\arcmin~16\arcsec$, in Aquila, 1.07~pc, 10.5~km~s$^{-1}$, and $(U, V, W)$=$(8.98, 4.92, -2.20)$~km~s$^{-1}$. For Voyager~1, 
         we obtained $\alpha=17^{\rm h}~31^{\rm m}~30^{\rm s}$ and $\delta=+12\degr~19\arcmin~2\arcsec$, in Ophiuchus, 1.70~pc, 16.7~km~s$^{-1}$, and 
         $(U, V, W)$=$(12.48, 8.85, 6.57)$~km~s$^{-1}$. For Voyager~2, we found $\alpha=21^{\rm h}~5^{\rm m}~3^{\rm s}$ and 
         $\delta=-67\degr~32\arcmin~29\arcsec$, in Pavo, 1.52~pc, 14.9~km~s$^{-1}$, and $(U, V, W)$=$(9.88, -6.49, -9.06)$~km~s$^{-1}$. If its current 
         trajectory does not change significantly, New Horizons will have $\alpha=19^{\rm h}~38^{\rm m}~54^{\rm s}$ and 
         $\delta=-19\degr~23\arcmin~23\arcsec$, in Sagittarius, 1.28~pc, 12.6~km~s$^{-1}$, and $(U, V, W)$=$(11.16, 4.14, -4.08)$~km~s$^{-1}$.
         For `Oumuamua, we found $\alpha=23^{\rm h}~51^{\rm m}~24^{\rm s}$ and $\delta=+24\degr~44\arcmin~59\arcsec$, in Pegasus, 2.69~pc,
         26.4~km~s$^{-1}$, and $(U, V, W)$=$(-5.89, 20.47, -15.56)$~km~s$^{-1}$. The corresponding values for 2I/Borisov from 
         \citet{2020MNRAS.495.2053D} are $\alpha=18^{\rm h}~21^{\rm m}~39^{\rm s}$ and $\delta=-51\degr~58\arcmin~37\arcsec$, in Telescopium, 1.65~pc, 
         32.28~km~s$^{-1}$, and $(U, V, W)$=$(29.50, -9.23, -9.30)$~km~s$^{-1}$. 

         The apexes of C/1980~E1 (Bowell) and C/2024~L5 (ATLAS) are located towards the same region of the sky (Fig.~\ref{apexes}, top panel) and away
         from those of 2I/Borisov and the probes. The future kinematics of C/1980~E1, C/2024~L5, and the probes is not too different from that of 
         stellar groups within 100~pc from the Sun as in table 1 of \citet{2016IAUS..314...21M} (Fig.~\ref{apexes}, bottom panels). On the other hand
         and as pointed out above, the receding velocities of C/1980~E1 and C/2024~L5 when entering interstellar space will be 3.8 and 2.8~km~s$^{-1}$, 
         respectively; in sharp contrast, those of interstellar probes will be significantly higher as shown above.

   \section{Conclusions\label{Con}}
      The Solar System can actively produce interstellar objects (see, e.g., \citealt{2018MNRAS.476L...1D}). Before 2024, only one such object was 
      known, C/1980~E1 (Bowell). Here, we studied its orbital evolution together with that of a recently discovered dynamical analog, C/2024~L5 
      (ATLAS), using direct $N$-body simulations and statistical analyses to explore the planetary encounters that led to their ejection from the 
      Solar System, and their pre- and post-encounter trajectories. Our conclusions can be summarized as follows.  
      \begin{enumerate}
         \item We confirm that C/1980~E1 reached its current path into interstellar space after an encounter with Jupiter at 0.23~au on December~9, 
               1980.
         \item We find that C/1980~E1 came from the inner Oort cloud.
         \item For C/1980~E1, we computed a receding velocity when entering interstellar space of 3.8~km~s$^{-1}$, moving towards Aries.
         \item We confirm that C/2024~L5 was scattered out of the Solar System following a flyby to Saturn at 0.003~au on January 24, 2022.
         \item We find that, prior to its planetary encounter, C/2024~L5 was probably a retrograde, inactive Centaur.
         \item The receding velocity of C/2024~L5 when entering interstellar space will be 2.8~km~s$^{-1}$, moving towards Triangulum.    
      \end{enumerate}
      Comet C/2024~L5 joins 1I/2017~U1, 2I/Borisov, and C/1980~E1 (Bowell) in the still small sample of known natural interstellar objects, although
      1I and 2I have a confirmed extrasolar provenance. Our results for two comets ejected from the Solar System confirm the widely accepted 
      conjecture that planetary systems such as the Solar System eject bodies throughout their lives and also provide constraints on the velocity 
      distribution of interstellar objects coming from dynamically evolved planetary systems.

   \begin{acknowledgements}
      We thank the anonymous referee for prompt, constructive, and helpful reports. We thank S. Deen for searching for precovery images of C/2024~L5 
      (ATLAS) and for comments on this particular object, and A.~I. G\'omez de Castro for providing access to computing facilities. Part of the 
      calculations and the data analysis were completed on the Brigit HPC server of the `Universidad Complutense de Madrid' (UCM), and we thank S. 
      Cano Als\'ua for his help during this stage. This research was partially supported by the Spanish `Agencia Estatal de Investigaci\'on 
      (Ministerio de Ciencia e Innovaci\'on)' under grant PID2020-116726RB-I00 /AEI/10.13039/501100011033. In preparation of this paper, we made use 
      of the NASA Astrophysics Data System, the ASTRO-PH e-print server, and the MPC data server.
   \end{acknowledgements}

   \bibliographystyle{aa}

\end{document}